# A Loophole in Bell's Theorem?
# Parameter Dependence in the Hess-Philipp Model

Wayne C. Myrvold[1]

Department of Philosophy

University of Western Ontario

London, ON Canada N6A 3K7

wmyrvold@uwo.ca

ABSTRACT

The hidden-variables model constructed by Karl Hess and Walter Philipp is claimed by its authors to exploit a "loophole" in Bell's theorem; according to Hess and Philipp, the parameters employed in their model extend beyond those considered by Bell. Furthermore, they claim that their model satisfies Einstein locality and is free of any "suspicion of spooky action at a distance." Both of these claims are false; the Hess-Philipp model achieves agreement with the quantum-mechanical predictions, not by circumventing Bell's theorem, but via Parameter Dependence.

**1. Introduction.** In recent papers, Karl Hess and Walter Philipp (2001*a,c*) exhibit a hidden-variables model that reproduces the quantum-mechanical statistics for the EPR-Bohm experiment. Of this model, they make the following claims:

1. The parameters used in the model go beyond those considered in the proofs of Bell's theorem, rendering the theorem inapplicable to their model.
2. The model satisfies the condition of Einstein locality and is free of "spooky action at a distance."

Both of these claims are false, as will be shown below.

**2. Factorization, Outcome Independence, Parameter Independence.** This section will introduce terminology and notation that will be used in subsequent sections. Much, if not all, of the material in this section will be familiar to many readers.

The EPR-Bohm experiment involves a pair of spin-½ particles prepared in the singlet state. Measurements are performed on each of the pair, at stations $S_1$ and $S_2$, respectively, of spin in directions **a** and **b**, respectively, where **a** and **b** are unit vectors. Denote by $A_\mathbf{a}$, $B_\mathbf{b}$ the outcomes, represented by values ±1, of the measurements on particle 1 and 2, respectively. The quantum-mechanical prediction for the expectation value of their product $A_\mathbf{a} B_\mathbf{b}$ is

$$E(\mathbf{a}, \mathbf{b}) = -\mathbf{a} \cdot \mathbf{b}.$$

Assume that the outcome of the experiment depends on a set of parameters denoted by $\lambda$. Without further assumptions about the nature of this dependence little can be said. The crucial assumption underlying the derivation of Bell's theorem is an assumption of factorizability of the underlying probability measures. Let $p_{\mathbf{ab}}(x_\mathbf{a}, y_\mathbf{b}| \lambda)$ be the probability that, in a state described by $\lambda$, the measurement of spin-**a** and spin-**b** on particles 1 and 2, respectively, have outcomes $x_\mathbf{a}$, $y_\mathbf{b}$, (the variables $x_\mathbf{a}$, $y_\mathbf{b}$ take on values in {-1, 1}). The Bell factorizability condition is the condition that there exist probability functions $p_\mathbf{a}(x_\mathbf{a}|\boldsymbol{\lambda})$, independent of **b**, and $p_\mathbf{b}(y_\mathbf{b}|\boldsymbol{\lambda})$, independent of **a**, such that

$$p_\mathbf{ab}(x_\mathbf{a}, y_\mathbf{b} | \boldsymbol{\lambda}) = p_\mathbf{a}(x_\mathbf{a} | \boldsymbol{\lambda}) \, p_\mathbf{b}(y_\mathbf{b} | \boldsymbol{\lambda}).$$

By constructing an inequality that must be satisfied by any factorizable theory but which is violated by the quantum-mechanical predictions, Bell was able to prove that no theory satisfying the Bell factorizability condition can reproduce the statistical predictions of quantum mechanics for the EPR-Bohm experiment. Bell's theorem is sometimes glossed as a proof of the impossibility of a hidden-variables theory, and this is how Hess and Philipp present it: "The work of Bell [(1964)] attempts to show that a mathematical description of EPR-type experiments by a statistical (hidden) parameter theory is not possible" (2001*c*, 14228). If this is what Bell had attempted to do, then his efforts would have been in vain—as Bell himself pointed out. In the very paper cited, Bell exhibits a simple hidden-variables model for the EPR-Bohm experiment, a model which is, of course, nonlocal. In his work, Bell frequently emphasized that it is not possible to show *tout court* that no hidden-variables theory is possible, as we actually have such a theory in Bohm's theory (Bohm 1952*a,b*). The paper (1966) that preceded (in order of composition, but not publication) the "Bell's Theorem" paper (1964) is devoted to criticizing purported no-go theorems; this includes the theorem that has come to be known as the Kochen-Specker theorem (Kochen and Specker, 1967), and which was proved, independently of Kochen and Specker, by Bell in his paper. Bohm's theory served Bell well as a counterexample to purported no-go theorems; a theorem purporting to show the impossibility of a hidden-variables theory must make some assumption not satisfied by the Bohm theory. In particular, the contextuality exhibited by the Bohm theory highlights the assumption of noncontextuality necessary for the Bell-Kochen-Specker theorem, and the nonlocality exhibited by the Bohm theory suggested the locality condition used to derive the Bell inequalities.

Jarrett (1984) showed that the Bell factorizability condition can be expressed as a conjunction of two conditions, the condition called "Locality" by Jarrett (1984; 1989) and "Parameter Independence" by Shimony (1986), on the one hand, and "Completeness" (Jarrett) or "Outcome Independence" (Shimony), on the other. Given the probability functions $p_{ab}(x_a, y_b|\lambda)$, we define marginal probability functions:

$$p_{ab}(x_a | \lambda) = \sum_{y_b \in \{-1, 1\}} p_{ab}(x_a, y_b | \lambda)$$
$$p_{ab}(y_b | \lambda) = \sum_{x_a \in \{-1, 1\}} p_{ab}(x_a, y_b | \lambda)$$

Parameter Independence is the condition that $p_{ab}(x_a|\lambda)$ not depend on **b** and $p_{ab}(y_b|\lambda)$ not depend on **a**. Outcome Independence is the condition that, for given **a**, **b**, $\lambda$, the distribution of $x_a$ is independent of $y_b$, and *vice versa*. That is,

$$p_{ab}(x_a, y_b | \lambda) = p_{ab}(x_a|\lambda) \, p_{ab}(y_b|\lambda).$$

A violation of Parameter Independence entails that, for a given value of $\lambda$, the statistical distribution of $x_a$ can be changed by changing the setting **b** of the distant analyzer. If the parameters $\lambda$ are observable, and if one can use them to adjust the settings of an analyzer, Parameter Dependence can be exploited to send a superluminal signal from one station to the other by manipulating the settings. If, for some reason, the parameters $\lambda$ cannot be used to adjust the settings of the analyzers, no superluminal signals can be sent but the nonlocality of the model remains, and, since the probability distribution of $x_\mathbf{a}$ depends on the *instantaneous* value of the setting at the other station, a model exhibiting Parameter Dependence requires a distinguished relation of distant simultaneity and cannot, at a fundamental level, be a relativistic model.

If we take, as the parameter $\lambda$, the quantum-mechanical state without any supplementary variables, the quantum-mechanical probabilities satisfy Parameter Independence but violate Outcome Independence. The peculiar sort of violation of Outcome Independence exhibited by quantum mechanics is, at the very least, not obviously at odds with relativity, as a violation of Parameter Independence would be (for discussion of this issue, see, *e.g.*, Shimony 1978, 1984, 1986; Jarrett 1984, 1989; Ballentine and Jarrett 1987; Butterfield 1989; Teller 1989; Redhead 1986, 1989; Myrvold 2002*a, b*).

**3. An Extended Parameter space?** According to Hess and Philipp, the basis for constructing a hidden-variables model that reproduces the quantum-mechanical statistical predictions lies in introducing time-dependent correlated parameters associated with the measurement instruments. Their construction proceeds via the introduction of setting-dependent subspace product measures (SDSPMs). Let our parameter space $\Omega$ be partitioned into disjoint subspaces $\Omega_m$. An SDSPM is a measure, defined on some subspace $\Omega_m$, whose dependence on the settings **a**, **b** is such that it can be written as a product of a measure dependent only on **a** and one dependent only on **b**. The probability measure constructed by Hess and Philipp is a sum of setting-dependent subspace product measures,

$$\mu_\mathbf{ab} = \frac{1}{N} \sum_{m=1}^{N} (\mu_\mathbf{a} \times \mu_\mathbf{b})_m .$$

Which of the product measures $(\mu_a \times \mu_b)_m$ is in effect on a given run of the experiment will be determined by the time-dependent correlated parameters.

Hess and Philipp assert that such measures go beyond those in the scope of Bell's theorem.

> We would like to emphasize that Bell has introduced a number of assumptions on time dependencies . . . He also introduced a significant asymmetry in describing the spin properties of the particles and the properties of the measuring equipment. The spins are described by arbitrarily large sets $\Lambda$ of parameters. On the other hand the measurement apparatus is described by a vector of Euclidean space (the settings), true to Bohr's postulate that the measurement must be classical. Yet, the measurement apparatus must itself in some form contain particles with spins that then, if one want to be self consistent, also need to be described by large sets of parameters that are related to the settings **a**, **b**, **c** . . . (2001c, 14229).

Now, the original proof of Bell's Theorem (Bell 1964) did not explicitly take into account hidden variables in the instruments. In his restatement of the proof (Bell 1971), Bell did explicitly include such a consideration.[2]

> The instruments themselves could contain hidden variables which could influence the results. If we average first over these instrument variables, we obtain the representation
> 
> $$P(\hat{a},\hat{b}) = \int d\lambda \, \rho(\lambda) \overline{A}(\hat{a},\lambda) \overline{B}(\hat{b},\lambda)$$

where the averages $\overline{A}$ and $\overline{B}$ will be independent of $\hat{b}$ and $\hat{a}$, respectively, if *the corresponding distributions of instrument variables are independent of b and a, respectively*, although of course they may depend on $\hat{a}$ and $\hat{b}$, respectively (1971, 178–79; 1987, 36–37).

The proof goes through under the assumption that the distribution of the instrument variables pertaining to one station is independent of the setting of the other (if this assumption is *not* made, it is of course possible to reproduce the statistical predictions of quantum mechanics and thereby violate the Bell Inequalities). The second of Hess and Philipp's claims, quoted above, is false; the derivation of the Bell inequalities in no way depends on an assumption that the states of the measurement instruments are completely described by their settings. This consideration also applies to the first above-quoted claim, that Bell introduces an assumption that the distributions of the hidden variables are time-independent; to generate time-dependent distributions, all one need do is include among the variables a parameter that changes with time in such a way that it can be used as a clock parameter. The key point is that, if correlations between events at the two wings of the experiment are due *solely* to clock parameters in the two instruments, then the probability of a result at one wing, conditional on a *full* specification of parameters, including clock parameters, at that wing, will be independent of the parameters at the other wing, even if the parameters pertaining to the two stations are correlated with each other.

Although Bell does not exclude the possibility of the measurement results depending on instrument parameters, he introduces a restriction on such parameters, namely, that the statistical distributions of parameters of one instrument be independent of the setting of the other instrument. It should be stressed that the locality condition assumed by Bell will be satisfied by clock parameters that are synchronized via some interaction in the common past of the instruments and subsequently evolve deterministically. If Hess and Philipp's construction is to exploit instrument parameters to yield the quantum-mechanical statistical predictions, and thereby violate the Bell inequality, it must have the probability distribution of the parameters for one instrument depend on the setting of the other—and it does. This feature of the Hess-Philipp model has been pointed out by Gill *et al.* (2002, 5).[3] In the next section a simplified version of Hess and Philipp's model will be exhibited, which will make the nature of the Parameter Dependence of such models clear.

**4. Parameter Dependence and Signalling**. Hess and Philipp claim that their model "is free of the suspicion of spooky action at a distance."

> This is accomplished by letting the probability measure be a superposition of setting-dependent subspace product measures with two important properties: (*i*) the factors of the product measure depend only on parameters of the station that they describe, and (*ii*) the joint density of the pairs of setting-dependent parameters in the two stations is uniform (2001*c*, 14232).

That is: the joint distribution of *u* and *v* is the uniform measure over a square in $\mathbb{R}^2$, and takes the form,

$$\mu_{\mathbf{ab}} = \frac{1}{N} \sum_{m=1}^{N} (\mu_{\mathbf{a}} \times \mu_{\mathbf{b}})_m .$$

The index *m*, which determines, for a given run of the experiment, which subspace measure will be in play, is a function of the source parameter $\lambda$ and the time parameters pertaining to the two stations. For each *m*, there are functions $A_m(\mathbf{a}, u)$, $B_m(\mathbf{b}, v)$ that determine the outcome of the experiments at 1 and 2, respectively.

Since $\mu_{\mathbf{ab}}$ is the same distribution for every **a**, **b**, changing the setting **b** will not affect the distribution, on $\mu_{\mathbf{ab}}$, of *u*, and changing the setting **a** will not affect the distribution of *v*. This does not, however, guarantee the absence of Parameter Dependence in the model.

Hess and Philipp's model is rather intricate, and there is not space to give the full details of the model here. Much of the intricacy, however, stems from the fact that they are concerned with reproducing the quantum-mechanical statistics for arbitrary settings **a**, **b** of the two analyzers. If we restrict the experimental set-up under consideration to one in which a choice is made between only two settings for each analyzer—a set-up which, of course, suffices for the violation of Bell inequalities— it is possible to construct a simpler model that nevertheless shares the salient features of Hess and Philipp's model, and, in particular, shares the features that are touted by Hess and Philipp as being responsible for the spooklessness of their model.

Let $a \in \{0, 1\}$ be a parameter indicating which of the two possible settings has been chosen for analyzer 1, and let $b \in \{0, 1\}$ indicate the setting of analyzer 2. Let *u* and *v* be hidden variables associated with analyzers 1 and 2, respectively, taking values in [0, 4). Define the functions $\sigma_a^i(u)$, $\tau_b^i(v)$, with values as given in Table 1.

|  | $u\in[0,1)$ | $u\in[1,2)$ | $u\in[2,3)$ | $u\in[3,4)$ |  | $v\in[0,1)$ | $v\in[1,2)$ | $v\in[2,3)$ | $v\in[3,4)$ |
|---|---|---|---|---|---|---|---|---|---|
| $\sigma_a^1(u)$ | $a$ | $1-a$ | $a$ | $1-a$ | $\tau_b^1(v)$ | $b$ | $b$ | $1-b$ | $1-b$ |
| $\sigma_a^2(u)$ | $1-a$ | $a$ | $1-a$ | $a$ | $\tau_b^2(v)$ | $b$ | $1-b$ | $1-b$ | $b$ |
| $\sigma_a^3(u)$ | $a$ | $1-a$ | $a$ | $1-a$ | $\tau_b^3(v)$ | $1-b$ | $1-b$ | $b$ | $b$ |
| $\sigma_a^4(u)$ | $1-a$ | $a$ | $1-a$ | $a$ | $\tau_b^4(v)$ | $1-b$ | $b$ | $b$ | $1-b$ |

**Table 1**

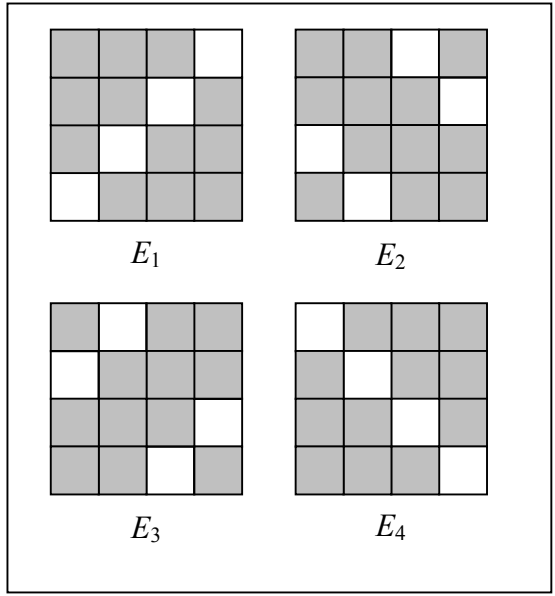

$E_1 \qquad E_2$

$E_3 \qquad E_4$

**Figure 1**

Let $E_j$, $j = 1,\ldots 4$, be the unshaded regions of the square $\Omega = [0, 4) \times [0, 4)$ depicted in Figure 1, and let $\kappa_j(u,v)$ be the function equal to 1 if $\langle u,v \rangle \in E_j$, and zero otherwise. Define

$$\rho_{ab}^{ij}(u,v) = \sigma_a^i(u)\,\tau_b^i(v)\,\kappa_j(u,v).$$

It is a simple matter to check that, for every value of $a$, $b$, and for each $i, j$,

$$\int_0^4 du \int_0^4 dv\, \rho_{ab}^{ij}(u,v) = 1.$$

$\rho_{ab}^{ij}(u,v)$ is, therefore, a probability density on $\Omega$. Let $\mu_{ab}^{ij}$ be the corresponding probability measures. Each $\rho_{ab}^{ij}$ is nonzero only on the subspace $E_j$, and, on that subspace, takes the form

$$\rho_{ab}^{ij}(u,v) = \sigma_a^i(u)\,\tau_b^i(v).$$

For each pair $i, j$, $\mu_{ab}^{ij}$ is therefore, a sum of setting-dependent subspace product measures. In what follows, we denote the pair $\langle i, j \rangle$ by the single index $m$, which takes on 16 values.

That fact that each $\rho_{ab}^m$ is, on the subspace picked out by $m$, a product of a function of $u$ and $a$ and a function of $v$ and $b$, might suggest that any correlations between $u$ and $v$ are due solely to the shared time parameter $m$ and that, in particular, the distribution of $u$ is independent of the distant setting $b$. This is not the case, because the subspace $E_j$ picked out by $m = \langle i, j \rangle$ is not a product space and has correlations between $u$ and $v$ built into it. In fact, each of the measures $\mu_{ab}^m$ exhibits a rather extreme form of Parameter Dependence. Note that, for each combination of $m, a, b$, the measure $\mu_{ab}^m$ assigns a nonzero probability to exactly one 1×1 square in $\Omega$, determined by the value of $a$ and $b$. Furthermore, for given $m$, if one knows *which* 1×1 square has nonzero probability, one can ascertain the values of $a$ and $b$. Thus, the dependence of the marginal distribution of $u$ on the value of $b$ is such that knowing the values, on a given run of the experiment, of $u$ (which must be a value assigned nonzero probability) and $m$ permits one to ascertain *both* $a$ and $b$. The same holds true of $v$. We therefore have Parameter Dependence—for given $m$, changing the value of $b$ affects the probability distributions for $u$, and changing the value of $a$ affects the probability distribution for $v$.

On the Hess-Philipp models, the corresponding Parameter Dependence is not quite as extreme as that exhibited by our simple model. The values of $m$ and $u$ on a particular run of the experiment, though they yield information about the setting of the distant apparatus, does not, on their model, permit one to ascertain precisely the value of the distant apparatus. The Parameter Dependence is, however, present (this can be ascertained by inspection of equation [44] of Hess and Philipp 2001a, or equation [27] of 2001c), and suffices for their model to reproduce the statistical predictions of quantum mechanics.

Given the fact that, on our model, knowledge of $m$, $u$ permits one to ascertain $b$, and knowledge of $m,v$ permits one to ascertain $a$, it would not be difficult, therefore, to cook up functions $A_m(a,u)$, $B_m(b,v)$ such that, for each $m$, the expectation value

$$\int du \int dv \, \rho_{ab}^m(u,v) \, A_m(a,u) B_m(b,v)$$

matches, for each of the four allowed combinations of the polarizers, the quantum mechanical expectation value $-\mathbf{a}\cdot\mathbf{b}$.

We will assume that associated with each measuring apparatus are synchronized time-dependent parameters, and that $m$ is a function of these, in such a way that, for any given run of the experiment, the same value of $m$ will be found at each station. We will also assume that, in any period of time long enough to include many runs of the experiment, $m$ ranges over all 16 of its permitted values, in such a way that, for each run of the experiment, each value of $m$ is equally probable. We will therefore consider the probability measure,

$$\mu_{ab} = \frac{1}{16} \sum_m \mu_{ab}^m$$

which has density function

$$\rho_{ab}(u,v) = \frac{1}{16} \sum_m \rho_{ab}^m(u,v).$$

It is easy to check that the measure $\mu_{ab}$ is uniform over the square $\Omega$. The fact that this is a uniform measure, however, has no bearing on the locality of the model; it is a sum of measures; each of which exhibits Parameter Dependence.

The Hess-Philipp model shares this feature with our simplified model. If the parameters *u*, *v*, and *m* are observable, and if it is possible to use the value of *m* to adjust the setting of the a measurement apparatus (and this is not excluded by anything they say), then signals can be sent from one station to the other by adjusting the settings of the other, because, if one knows the value of *m* for each run of the experiment (a value that is determined by correlated parameters at each station), observation of the statistical distribution of *u* at station 1 yields information about the setting **b** at station 2. If the parameters *u*, *v*, and *m* are not observable, this obscures the nonlocality in the model but does not remove it; the dependence of the distribution of parameters associated with one station on the setting of the other remains an inherent feature of the model. The Hess-Philipp model achieves agreement with the quantum-mechanical predictions, not by introducing parameters of a sort left out of consideration by Bell, but by using parameters of a sort considered by Bell and violating the locality conditions assumed by Bell.

**Notes**

[1] I am grateful to Jim Brown for encouraging me to look at Hess and Philipp's paper, and to Richard Gill for helpful discussions. Since the completion of the draft of the version of this paper originally submitted for presentation at PSA, two other criticisms of Hess and Philipp have been put forward, that of Gill *et al.* (2002), and that of Mermin (2002). There is agreement among these criticisms that Hess and Philipp have not shown that there is a flaw in the standard proofs of Bell's theorem, and that their model achieves its result via nonlocality. It is hoped that the present paper is a useful supplement to the other criticisms and that, in particular, the simplified model of section 4 helps to make it clearer what Hess and Philipp *have* done.

[2] In a subsequent paper, Hess and Philipp remark that "Bell has included into his later proofs (after publication of [Ballentine and Jarrett 1987]) setting-dependent parameters" (Hess and Philipp 2002, 780). This remark, while true, is doubly misleading. First, it suggests that Bell did not include such considerations prior to 1987. Second, it suggests that the cited paper is in some way relevant to the consideration of such parameters. Ballentine and Jarrett's excellent paper is, however, about another topic entirely, namely, the distinction between "completeness" and "locality" mentioned in section 2.

[3] In a paper published subsequently to the papers presenting the model under discussion here, Hess and Philipp remark, "Of course, to obey Einstein locality, [the station parameters] must be station specific and can only be correlated by time-like correlations, *i.e.* by some relation to local periodic processes" (Hess and Philipp 2002, 780). This is a clear statement of what their model fails to do.

18